\documentclass{ws-p10x7_clns}

\begin{document}

\title{First Observation of the \lowercase{$\Sigma_c^{\ast +}$}
Charmed Baryon, and New Measurements of the \lowercase{$\Sigma_c^0$},
\lowercase{$\Sigma_c^+$}, \lowercase{$\Sigma_c^{++}$}, and
\lowercase{$\Omega_c^0$} Charmed Baryons}

\author{Andreas Warburton\\(representing the CLEO Collaboration)}

\address{Laboratory of Nuclear Studies, Cornell University,
Ithaca, New York 14853, USA\\E-mail: andreas.warburton@cornell.edu}

\twocolumn[\maketitle
\abstract{ Using $\sim$13.7~fb$^{-1}$ of
$e^+\,e^-$ collision data recorded at near the $\Upsilon(4S)$
resonance by the CLEO~II and CLEO~II.V detector configurations on the
Cornell Electron Storage Ring, we present the world's most precise
measurements of the $\Sigma_c^0$, $\Sigma_c^+$, and $\Sigma_c^{++}$
charmed-baryon masses as well as the first measurements of the
intrinsic widths of the $\Sigma_c^0$ and $\Sigma_c^{++}$ baryons.  We
also report on the first observation and mass measurement of the
$\Sigma_c^{\ast +}$ charmed baryon, $M(\Sigma_c^{\ast +}) -
M(\Lambda_c^+) = (231.0\pm 1.1[{\rm stat}]\pm 2.0[{\rm
syst}])$~MeV/$c^2$, and the first CLEO observation of the $\Omega_c^0$
baryon, for which we measure a mass $M(\Omega_c^0) = (2694.6\pm
2.6[{\rm stat}]\pm 1.9[{\rm syst}])$~MeV/$c^2$ from a sample of
$(40.4\pm 9.0[{\rm stat}])$ candidate events.  All new results are
preliminary.}]

\section{Introduction}
The charmed baryons consist of a $c$ quark and a light diquark with a
specific J$^{\rm P}$ spin-parity configuration and can be organized in
terms of isospin and strangeness using the theory of SU(4) multiplets.
Several new observations and improved measurements in charmed-baryon
spectroscopy have been made in the past decade\footnote{Note, however,
that none of the J$^{\rm P}$ quantum numbers of the charmed baryons
has yet been measured.}.  We report here on new CLEO measurements of
the $\Sigma_c$ and $\Omega_c^0$
states~\cite{cleo_sigmacsp,cleo_omegac}.

\section{The Experiment}
\label{sect:expt}
The results described herein come from studies of $e^+\,e^-$
collisions conducted at the Cornell Electron Storage Ring (CESR)
operating near the $\sim$10.58~GeV/$c^2$ $\Upsilon(4S)$ bottomonium
resonance.  The $4\pi$ general purpose CLEO~II detector
configuration~\cite{cleo_nim}, comprising a cylindrical drift chamber
system for charged-track detection inside a 1.4~T solenoidal magnetic
field and a CsI electromagnetic calorimeter for $\pi^0$ detection, was
employed to take a data sample with time-integrated luminosity
$\int\!\!{\cal L}dt \simeq 4.7$~fb$^{-1}$.  An additional
$9.0$~fb$^{-1}$ of data were taken with the CLEO~II.V detector
configuration~\cite{cleo_nim_svx}, which had improved charged-track
measurement capabilities.

For the searches and measurements described in this paper, the
analysis approaches have sought to optimize signal efficiency and
background suppression.  Identification of $p$, $K^+$, and $\pi^+$
candidates\footnote{Charge conjugate modes are implicit throughout.}
was achieved through the use of specific ionization $dE/dx$ in the
drift chamber and, when available, time-of-flight information.
Hyperons, in the modes $\Xi^- \to \Lambda\,\pi^-$, $\Xi^0 \to
\Lambda\,\pi^0$, $\Omega^- \to \Lambda\, K^-$, $\Sigma^+\to p\,\pi^0$,
and $\Lambda\to p\,\pi^-$, were reconstructed by detecting their decay
points separated from the primary event vertex.

Charmed baryons at CESR are either produced from the secondary decays
of $B$ mesons or directly from $e^+\,e^-$ annihilation to $c\bar{c}$
jets.  Combinatorial backgrounds, which are highest for low-momentum
charmed baryons, are reduced by requiring candidates to exceed an
optimum scaled momentum $x_p$, defined as $x_p \equiv
\left|\vec{p}\right| / p_{\rm max}$, where $\vec{p}$ is the momentum
of the charmed baryon candidate, $p_{\rm max} \equiv \sqrt{E^2_{\rm
beam} - M^2}$, $E_{\rm beam}$ is the beam energy, and $M$ is the
reconstructed candidate's mass.  We use decay-mode dependent
scaled-momentum criteria\footnote{Charmed baryons with
threshold-produced $B$-meson parents are kinematically limited to $x_p
< 0.4$.} of $x_p > 0.5$ or $x_p > 0.6$, based on the optimization
results.

\section{The $\Sigma_c$ and $\Sigma_c^\ast$ Baryon Isotriplets}
\label{sect:sigmac}
The $\Sigma_c$ baryons~\cite{sigc_notation} consist of one charm and
two light ($u$ or $d$) valence quarks in an $I=1$ isospin
configuration.  The $\Sigma_c$ states in this study decay strongly
into final states with a $\Lambda_c^+$ baryon and either a $\pi^+$ or
$\pi^0$ transition meson.  We reconstruct $\sim$58$\, 000$
$\Lambda_c^+$ signal candidates using 15 different decay
modes~\cite{cleo_lambdac}.

\subsection{$\Sigma_c$ Final States involving $\Lambda_c^+\,\pi^\pm$}
\label{sect:sigc_pip}
\begin{figure}
\epsfxsize170pt
\figurebox{120pt}{160pt}{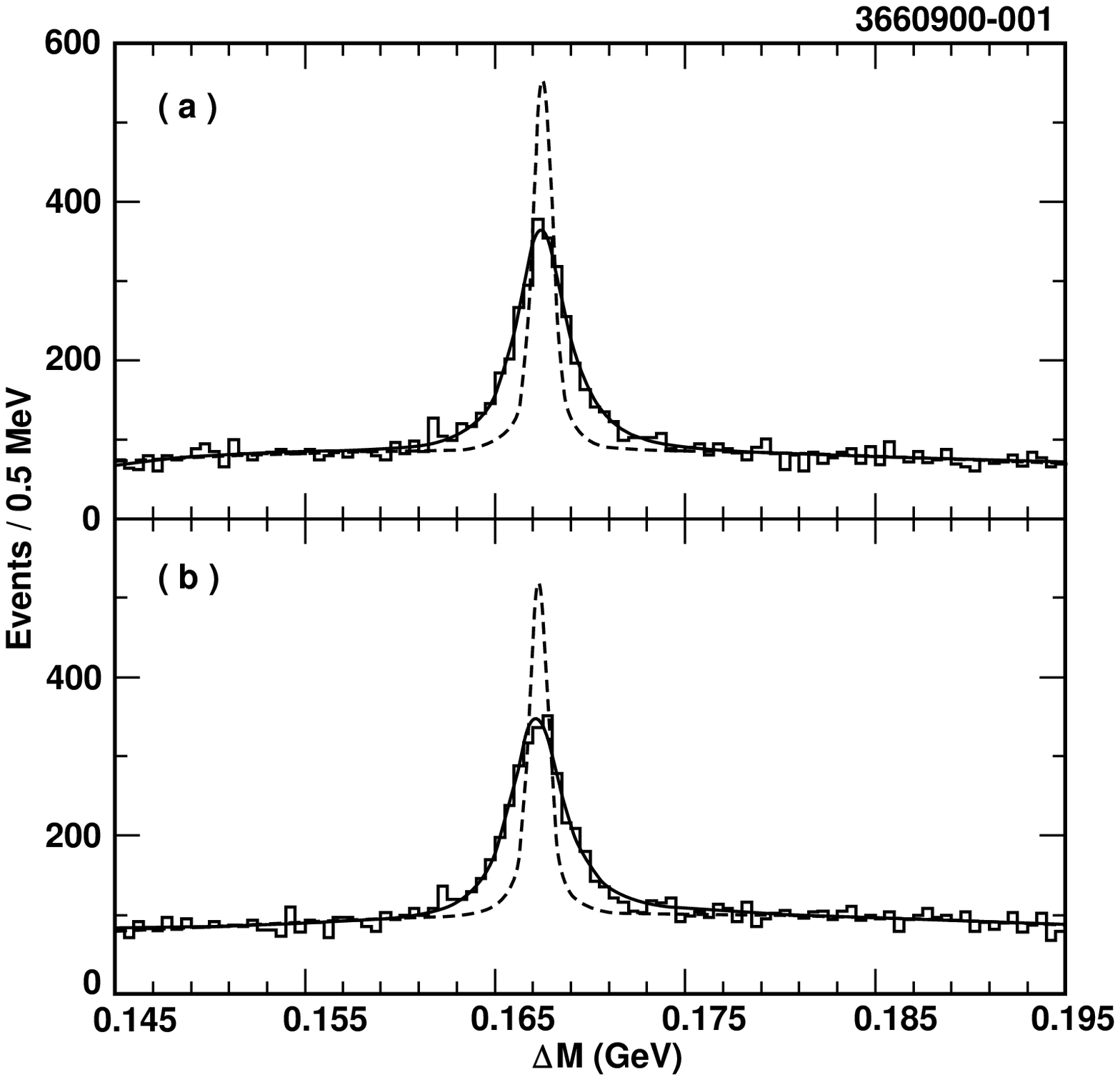}
\caption{The $\Delta M \equiv M(\Lambda_c^+\,\pi^\pm) -
M(\Lambda_c^+)$ mass difference, where in (a) the $\Lambda_c$ and
$\pi$ candidates have like charges ($\Sigma_c^{++}$ candidates) and in
(b) they have opposite charges ($\Sigma_c^{0}$ candidates).  The solid
curve shows the results of the fit described in the text.  The dashed
curve represents the resolution function of the detector response.}
\label{fig:sigc}
\end{figure}

Candidates in the $\Lambda_c^+$ sample described in
Sec.~\ref{sect:sigmac} were combined with $\pi^\pm$ charged-track
candidates and the mass difference, $\Delta M \equiv
M(\Lambda_c^+\,\pi^\pm) - M(\Lambda_c^+)$, was calculated for those
combinations satisfying an $x_p > 0.5$ criterion.  Each of the
resulting mass-difference distributions, depicted in
Fig.~\ref{fig:sigc}, indicated an unambiguous signal of $\sim$2000
$\Sigma_c^{++}$ and $\Sigma_c^0$ candidates, respectively.  The two
distributions were fit to a sum of a polynomial background function
with threshold suppression and a $p$-wave Breit-Wigner line shape
convolved with a double-Gaussian detector resolution function.

\begin{table*}
\begin{center}
\caption{A summary of the CLEO mass and width measurements, in
MeV/$c^2$, of the ${\rm J}^{\rm P} = \frac{1}{2}^+$ and ${\rm J}^{\rm
P} = \frac{3}{2}^+$ $\Sigma_c$ charmed baryons.  The first
uncertainties are statistical and the second are systematic.  Values
marked with a $\dagger$ were reported in
Ref.~\protect\citelow{cleo_sigmaold}.  We note that the isospin
mass splittings are consistent with theoretical
expectation~\protect\citelow{franklin}.}\label{tab:sigc}
\begin{tabular}{|c|c|c|c|c|}
 
\hline
${\rm J}^{\rm P}$ & Observable & $\Sigma_c^{(\ast)0}$ &
$\Sigma_c^{(\ast)+}$ & $\Sigma_c^{(\ast)++}$ \\ \hline\hline
 & $M(\Sigma_c) - M(\Lambda_c^+)$ & $167.2\pm 0.1\pm 0.2$ &
                                 $166.4\pm 0.2\pm 0.3$ &
                                 $167.4\pm 0.1\pm 0.2$ \\
\raisebox{6pt}[0pt][0pt]{${1 \over 2}^+$}
                                 & $\Gamma(\Sigma_c)$ &
                                 $2.4\pm 0.2\pm 0.4$   &
                                 $< 4.6\ (90\%\ {\rm CL})$&
                                 $2.5\pm 0.2\pm 0.4$ \\ \hline
 & $M(\Sigma_c^\ast) - M(\Lambda_c^+)$ & $232.6\pm 1.0\pm 0.8^\dagger$ &
                                      $231.0\pm 1.1\pm 2.0$ &
                                      $234.5\pm 1.1\pm 0.8^\dagger$ \\
\raisebox{6pt}[0pt][0pt]{${3 \over 2}^+$}
                                 & $\Gamma(\Sigma_c^\ast)$ &
                                 $13.0\, ^{+3.7}_{-3.0}\, \pm 4.0^\dagger$ &
                                 $< 17\ (90\%\ {\rm CL})$           &
                                 $17.9\, ^{+3.8}_{-3.2}\, \pm 4.0^\dagger$ \\
\hline
\end{tabular}
\end{center}
\end{table*}

The fitted masses and widths are summarized in Tab.~\ref{tab:sigc}.
Scaling from the strange-baryon widths, Rosner~\cite{rosner} has
predicted $\Gamma(\Sigma_c) \simeq 1.32$~MeV/$c^2$, somewhat narrower
than our measured values.  Other authors~\cite{sigc_theory} have
employed measurements of the $\Sigma_c^\ast$ widths to derive
$\Gamma(\Sigma_c)$ predictions consistent with our values listed in
Tab.~\ref{tab:sigc}.

\subsection{$\Sigma_c^{(\ast)}$ Final States involving $\Lambda_c^+\,\pi^0$}
\begin{figure}
\epsfxsize170pt
\figurebox{120pt}{160pt}{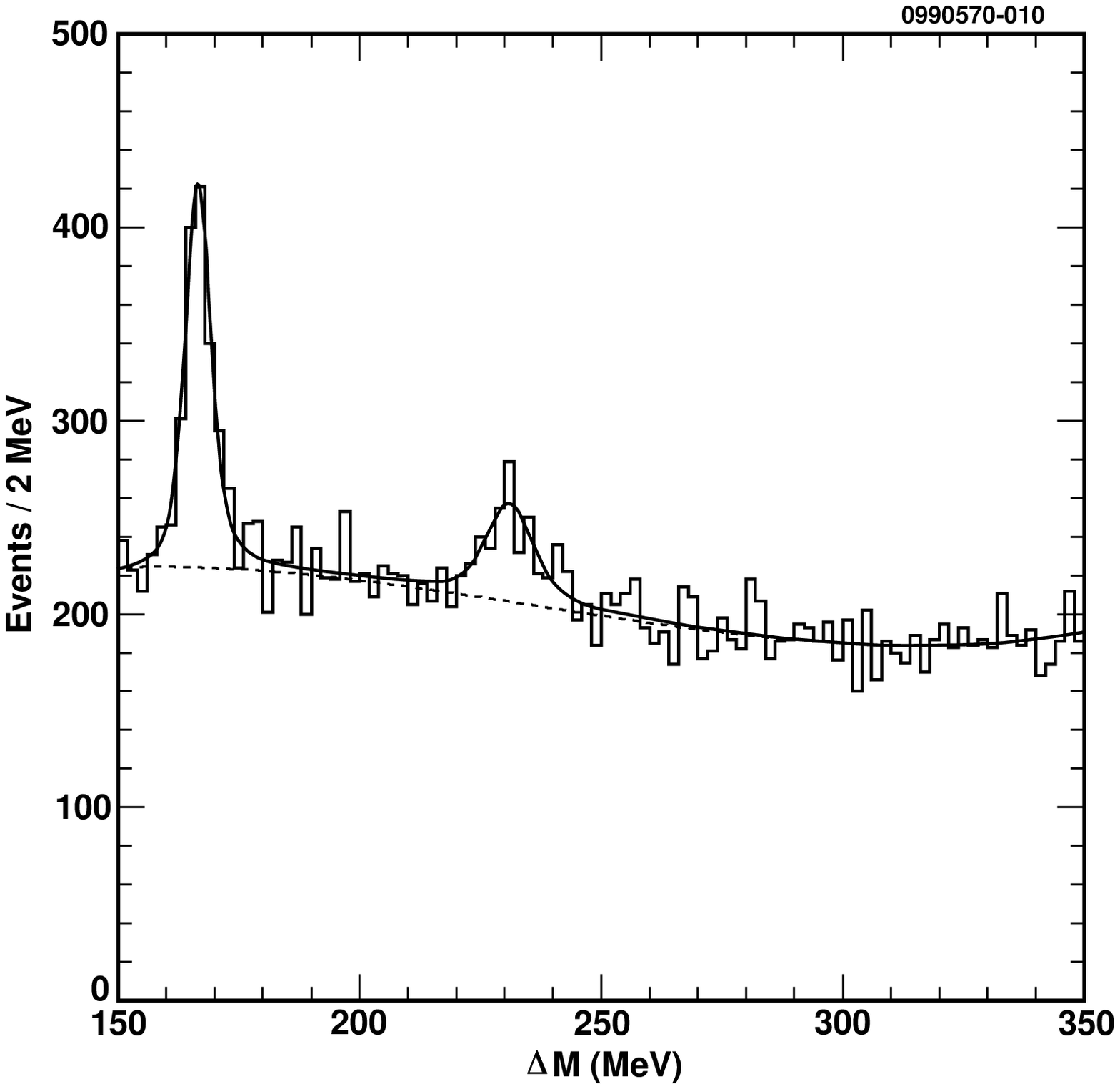}
\caption{The $\Delta M \equiv M(\Lambda_c^+\,\pi^0) - M(\Lambda_c^+)$
mass-difference distribution.  The solid line represents a fit to a
third-order polynomial background shape and two $p$-wave Breit-Wigner
functions smeared by Gaussian resolution functions for the two signal
shapes.  The dashed curve indicates the background function.}
\label{fig:sigcstplus}
\end{figure}

In a manner similar to that described in Sec.~\ref{sect:sigc_pip} but
with an $x_p > 0.6$ requirement~\cite{cleo_sigmacsp}, $\Lambda_c^+$
candidates were combined with transition $\pi^0$ candidates, which
were required to have momenta greater than 150~MeV/$c$ and masses
consistent with and kinematically fit to the known $\pi^0$
mass~\cite{PDG}, to form the mass difference $\Delta M \equiv
M(\Lambda_c^+\,\pi^0) - M(\Lambda_c^+)$ shown in
Fig.~\ref{fig:sigcstplus}.  Note the higher background level.

The lower-mass peak\footnote{Note that, for the purpose of yield
determination, the fit function described in Fig.~\ref{fig:sigcstplus}
was replaced with a third-order Chebyshev polynomial background
function and two Gaussian signal line shapes.} in
Fig.~\ref{fig:sigcstplus}, containing $(661^{+63}_{-60}[{\rm stat}])$
events, is due to the $\Sigma_c^+$ baryon.  The second peak in
Fig.~\ref{fig:sigcstplus}, with $(327^{+78}_{-73}[{\rm stat}])$ signal
candidates, we ascribe to the first observation of $\Sigma_c^{\ast +}$
baryons.  The fitted masses and limits on the intrinsic widths of the
$\Sigma_c^+$ and $\Sigma_c^{\ast +}$ states are listed in
Tab.~\ref{tab:sigc}.

\begin{figure}
\epsfxsize170pt
\figurebox{120pt}{160pt}{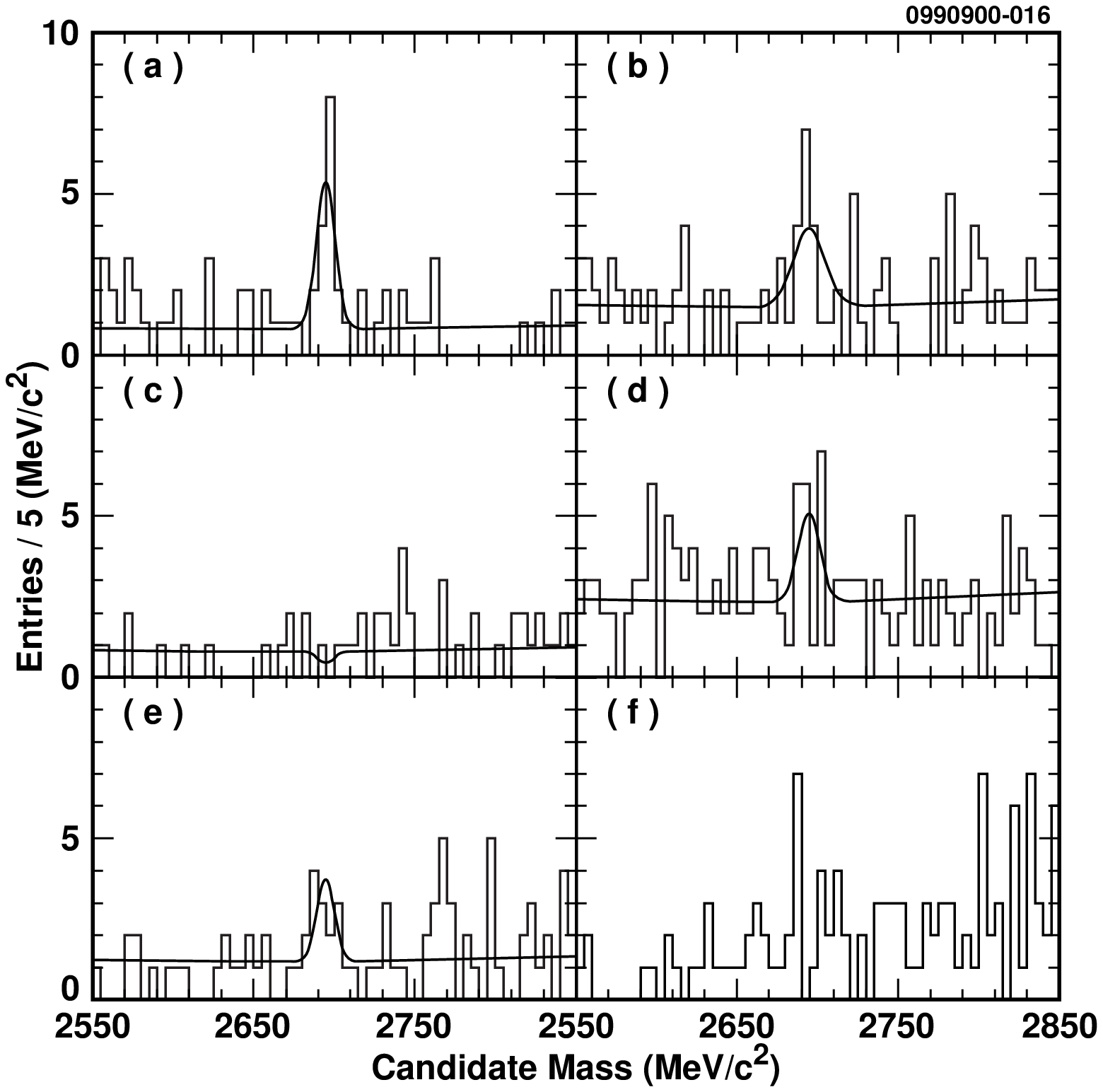}
\caption{The invariant mass distribution and simultaneous fit to the
five $\Omega_c^0$ search modes: (a) $\Omega^-\,\pi^+$, (b)
$\Omega^-\,\pi^+\,\pi^0$, (c) $\Omega^-\,\pi^+\,\pi^+\,\pi^-$, (d)
$\Xi^0\, K^-\,\pi^+$, and (e) $\Xi^-\, K^-\,\pi^+\,\pi^+$.  The final
state (f) $\Sigma^+\, K^-\, K^-\, \pi^+$ was not included in the fit.
The signal region was fitted with a fixed-width Gaussian while the
background was fitted to a second-order polynomial.}
\label{fig:omegac_modes}
\end{figure}

\section{The $\Omega_c^0$ Baryon Isosinglet}
The $\Omega_c^0$ baryon is a J$^{\rm P} = \frac{1}{2}^+$ ground state
with valence quarks $(c\{ss\})$, where $\{ss\}$ denotes a symmetry in
the wave function under the exchange of the light-quark spins.
Several experimental groups have reported the observation of an
$\Omega_c^0$ state; the mutual consistency of the claimed masses,
however, is marginal~\cite{cleo_omegac,PDG}.

\subsection{First $\Omega_c^0$ Observation by CLEO}
\begin{figure}
\epsfxsize170pt
\figurebox{120pt}{160pt}{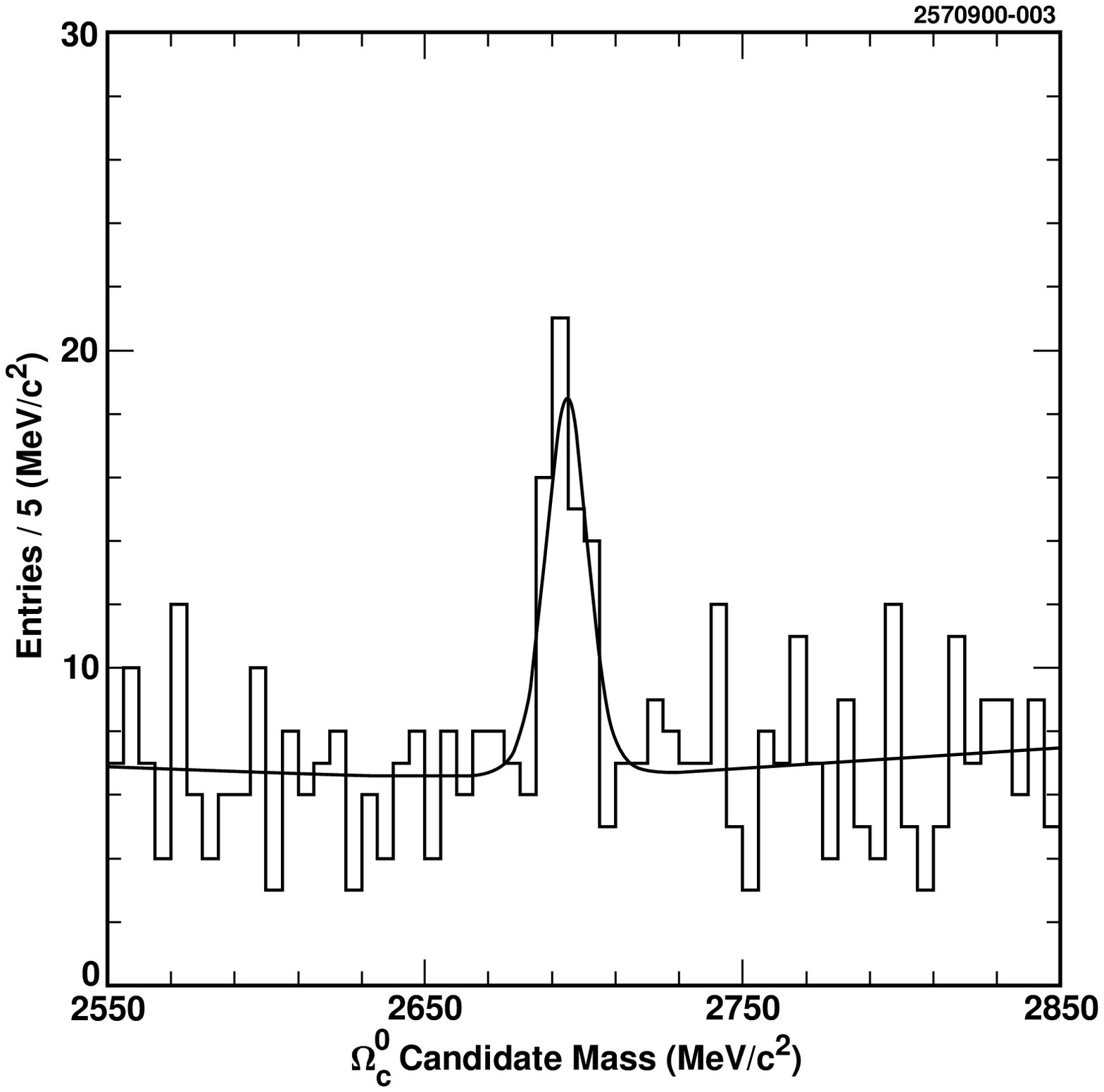}
\caption{The invariant mass distribution (histogram) for the sum of
the $\Omega^-\,\pi^+$, $\Omega^-\,\pi^+\,\pi^0$,
$\Omega^-\,\pi^+\,\pi^+\,\pi^-$, $\Xi^0\, K^-\,\pi^+$, and $\Xi^-\,
K^-\,\pi^+\,\pi^+$ search modes.  The fit function (solid curve) is
the sum of the fit functions indicated in Fig.~\ref{fig:omegac_modes}.}
\label{fig:omegac_total}
\end{figure}

Based on patterns observed in other charmed baryon decays and
considerations of reconstruction efficiency and combinatorial
background, we searched~\cite{cleo_omegac} for $\Omega_c^0$ candidates
in the five weak decay modes $\Omega^-\,\pi^+$,
$\Omega^-\,\pi^+\,\pi^0$, $\Omega^-\,\pi^+\,\pi^+\,\pi^-$, $\Xi^0\,
K^-\,\pi^+$, and $\Xi^-\, K^-\,\pi^+\,\pi^+$.  We separately
investigated a sixth channel, $\Sigma^+\, K^-\, K^-\, \pi^+$, because
E687~\cite{cleo_omegac,PDG} showed a significant signal in this mode.
The invariant mass distributions of the six modes are shown in
Fig.~\ref{fig:omegac_modes}, the sum of the five search modes is
depicted in Fig.~\ref{fig:omegac_total}, and the fitted event yields
are listed in Tab.~\ref{tab:omegc}.  We observe $(40.4\pm 9.0[{\rm
stat}])$ $\Omega_c^0$ candidates.

\begin{table*}
\begin{center}
\caption{The $\Omega_c^0$ charmed baryon search results in the six
decay modes.  The fitted yields were computed with a mode-dependent
$x_p$ criterion, whereas the relative branching fraction (${\cal B}$)
and cross-section-branching-fraction product ($\sigma\cdot {\cal B}$)
were determined with a uniform $x_p > 0.5$ requirement.  The first
uncertainties are statistical and the second are
systematic.}\label{tab:omegc}
\begin{tabular}{|c|c|c|c|} 
 
\hline
$\Omega_c^0$ Search Channel & Fitted Yield & Relative ${\cal B}$
	& $\sigma \cdot {\cal B}$~[fb] \\
\hline\hline
$\Omega^-\,\pi^+$
	& $13.3\pm 4.1$ & $1.0$ & $11.3\pm \phantom{0}3.9\pm 2.0$ \\
$\Omega^-\,\pi^+\,\pi^0$
	& $11.8\pm 4.9$ & $4.2\pm 2.2\pm 0.9$ & $47.6\pm 18.0\pm 3.1$\\
$\Xi^0\, K^-\,\pi^+$
	& $\phantom{0}9.2\pm 4.9$
                        & $4.0\pm 2.5\pm 0.4$ & $45.1\pm 23.2\pm 3.7$\\
$\Xi^-\, K^-\,\pi^+\,\pi^+$
	& $\phantom{0}7.0\pm 3.7$
                        & $1.6\pm 1.1\pm 0.4$ & $18.2\pm 10.6\pm 3.3$\\
$\Omega^-\,\pi^+\,\pi^+\,\pi^-$
	& $-0.9\pm 1.4\phantom{\,}$
                        & $< 0.56$~(90\% CL)  & $< 5.1$~(90\% CL) \\
\hline
Sum of the 5 modes above
	& $40.4\pm 9.0$ & $-$                 & $-$ \\
\hline
$\Sigma^+\, K^-\, K^-\, \pi^+$
	& $\phantom{0}2.8\pm 4.1$
                        & $< 4.8$~(90\% CL)   & $< 53.8$~(90\% CL) \\
\hline
\end{tabular}
\end{center}
\end{table*}

\subsection{Measurement of the $\Omega_c^0$ Mass}
The mass of our $\Omega_c^0$ candidates we measure by performing an
unbinned maximum-likelihood fit using the sum of a single Gaussian
signal and a second-order polynomial background.  We find the
$\Omega_c^0$ mass to be $M(\Omega_c^0) = (2694.6\pm 2.6[{\rm stat}]\pm
1.9[{\rm syst}])$~MeV/$c^2$, where the systematic uncertainty is
dominated by our sensitivity to the fitting method employed.

\section{Conclusion}
The CLEO collaboration has made new measurements of the $\Sigma_c^0$,
$\Sigma_c^+$, and $\Sigma_c^{++}$ baryon masses as well as preliminary
measurements of the $\Sigma_c^0$ and $\Sigma_c^{++}$ intrinsic widths.
We report the first observation of the $\Sigma_c^{\ast +}$ baryon and
determine its mass difference to be $M(\Sigma_c^{\ast +}) -
M(\Lambda_c^+) = (231.0\pm 1.1[{\rm stat}]\pm 2.0[{\rm
syst}])$~MeV/$c^2$.  We also observe, for the first time, a
significant signal for the $\Omega_c^0$ baryon and measure its mass to
be $M(\Omega_c^0) = (2694.6\pm 2.6[{\rm stat}]\pm 1.9[{\rm
syst}])$~MeV/$c^2$.

\section*{Acknowledgments}
\vspace{-0.125in}
My colleagues in the CLEO collaboration, the staff at CESR, and their
funding sources made these results possible.  I thank Basit Athar and
John Yelton for useful discussions.


\vspace{0.2in}
\begin{thebibliography}{99}
\vspace{-0.2in}
\bibitem{cleo_sigmacsp}
R.~Ammar {\it et al.} (CLEO),~[hep-ex/000\\ 7041], (submitted to {\it
Phys.\ Rev.\ Lett.}).

\bibitem{cleo_omegac}
D.~Cronin-Hennessy~{\it et al.}~(CLEO),~[hep-ex/0010035], (submitted
to {\it Phys.\ Rev.\ Lett.}), and citations therein.

\bibitem{cleo_nim}
Y. Kubota {\it et al.} (CLEO), {\it Nucl. Instrum. Methods}
{\bf A320}, 66 (1992).

\bibitem{cleo_nim_svx}
T.~S. Hill, {\it Nucl. Instrum. Methods} {\bf A418}, 32 (1998).

\bibitem{sigc_notation}
The symbols $\Sigma_c$ and $\Sigma_c^\ast$ refer to the
$\Sigma_c (2455)$ and $\Sigma_c (2520)$ baryons, respectively.

\bibitem{cleo_lambdac}
P.\ Avery {\it et al.} (CLEO), {\it Phys.\ Rev.\ D} {\bf 43}, 3599 (1991);
{\it idem}, {\it Phys.\ Rev.\ Lett.} {\bf 71}, 2391 (1993);
{\it idem}, {\it Phys.\ Lett.} {\bf B325}, 257 (1994);
M.~S.\ Alam {\it et al.}, {\it Phys.\ Rev.\ D} {\bf 57}, 4467 (1998),
[hep-ex/9709012].

\bibitem{cleo_sigmaold}
G. Brandenburg {\it et al.} (CLEO), {\it Phys. Rev. Lett.}
{\bf 78}, 2304 (1997).

\bibitem{franklin}
J. Franklin, {\it Phys. Rev. D} {\bf 59}, 117502 (1999), [hep-ph/9901294].

\bibitem{rosner}
J.~L. Rosner, {\it Phys.\ Rev.\ D} {\bf 52}, 6461 (1995), [hep-ph/9508252].

\bibitem{sigc_theory}
D. Pirjol and T.-M. Yan, {\it Phys.\ Rev.\ D} {\bf 56}, 5483 (1997),
[hep-ph/9701291];
S.~Tawfiq {\it et al.}, {\it Phys.\ Rev.\ D} {\bf 58}, 054010 (1998),
[hep-ph/9803246];
M.~A. Ivanov {\it et al.}, {\it Phys.\ Lett.\ B} {\bf 442}, 435 (1998),
[hep-ph/9807519].

\bibitem{PDG}
D.~E. Groom {\it et al.} (PDG), {\it Eur. Phys. J. C}
{\bf 15}, 1 (2000).


\end{thebibliography}
\end{document}